\documentclass[12pt]{article}
\usepackage[cp850]{inputenc}
\usepackage{amsmath,amsfonts,amssymb,amsthm}
\usepackage[english]{babel}
\usepackage{latexsym}
\usepackage{amscd}
\usepackage{float}
\newtheorem{thm}{Theorem}

\DeclareMathOperator{\HH}{H}
\DeclareMathOperator{\D}{D}

\begin{document}
\title{Interaction between Truth and Belief as the key to entropy and
other quantities of statistical physics}

\author{
  Flemming Tops{\o}e \\  University of Copenhagen, Department of
Mathematical Sciences\\
Universitetsparken 5, 2100 Copenhagen, Denmark\\topsoe@math.ku.dk}
\date{}

\maketitle
\begin{abstract}
  
  The notion of {\it entropy} penetrates much of science. A key
  feature of the all-important notion of {\it Boltzmann-Gibbs-Shannon
    entropy} is its {\it extensivity} (additivity over independent
  subsystems). However, there is a need for other quantities. In
  statistical physics a parameterized family of {\it non-extensive}
  entropy measures, now mainly known under the name of {\it
    Tsallis-entropies}, have received much attention but also been met
  with criticism due mainly to a lack of convincing interpretations. 

  Based on the hypothesis that interaction between {\it truth}, held
  by ``Nature'', and {\it belief}, as expressed by man, may take
  place, classical- as well as non-classical measures of entropy and
  other essential quantities are derived. The approach aims at
  providing a genuine interpretation, rather than relying either on
  analogies based on formal mathematical manipulations or else -- 
  more fruitfully, but not satisfactory
  -- on axiomatic characterizations.
    
\end{abstract}

\section{Contemplation}

Let us apply a philosophical approach and put ourselves in the shoes
of the physicist, planning to set-up {\it experiments} and to engage
in associated {\it observations}. He might argue as follows: 
\medskip

\medskip {\bf 1:} As an expression of my {\it beliefs} concerning phenomena I
plan to observe, I shall assign numbers in $[0,1]$, typically denoted
by the letter $y$, to {\it events} associated with the phenomena.
Certain, to me unknown numbers, likewise in $[0,1]$, and likewise
associated with events, 
express the essence of the phenomena, and do not depend on my
interference. They are referred to as {\it truth-assignments} and are,
typically, denoted by the letter $x$.

\medskip {\bf 2:} Any event I may observe entails a certain {\it effort} on my
part. This effort I shall also refer to as {\it individual complexity}
-- ``individual'', because it is associated with each individual event
I could encounter.  Before setting up {\it experiments}, I should
determine the effort I am willing to or have to devote to any event I
may be faced with. This should depend only on the assigned
belief-value $y$, and is denoted $\kappa(y)$. The function $\kappa$,
defined on $[0,1]$ and with values in $[0,\infty]$, I refer to as the
{\it coder}. As $1$ represents {\it certainty}, I insist that
$\kappa(1)=0$.  Further, I assume that $\kappa$ is smooth in a
technical sense, say continuous on $[0,1]$ and continuously
differentiable and finite valued on $]0,1]$. Finally,
as I do not want to distinguish between coders that only differ by a
scalar factor, I will introduce an
assumption of {\it normalization} in order to pick out a canonical
representative of the possible coders. As $\kappa(1)=0$ and as I do not
want to assume that $\kappa(0)$ is finite, I choose to impose the
condition $\kappa'(1)=-1$ for the stated purpose.
      
\medskip {\bf 3} To determine the coder, I must know the basic characteristica
of the {\it world} I operate in. I choose to focus primarily on a
concept of {\it interaction between truth and belief}.

\medskip {\bf 4}  I shall model this interaction by a function
$\pi$ defined on the product set $[0,1]\times[0,1]$ and taking values
in $]-\infty,\infty]$. The idea is that
$\pi(x,y)$ represents the {\it force} by which the world presents an
event to me in case the truth-assignment is $x$ and my belief
in the event is $y$. On the technical side, I better assume that $\pi$
is continuous on its domain and continuously differentiable and
finite-valued on $[0,1]\times]0,1]$.

\medskip {\bf 5} I consider the {\it classical world} to be a world of
{\it ``no interaction''}, i.e. $\pi(x,y)=x$ for all $(x,y)$. I must be
prepared for other forms of interaction, but will always assume that
the interaction is {\it sound}, i.e. that $\pi(x,x)=x$ for all
$x\in[0,1]$. Stronger conditions should be considered and in this
connection, it appears sensible to impose conditions of {\it
  consistency}: I will call the interaction {\it weakly consistent} if
$\sum_{i\in\mathbb A}\pi(x_i,y_i)=1$ for any finite {\it alphabet}
$\mathbb A$ and any truth-assignment $x=(x_i)_{i\in\mathbb A}$ and
belief-assignment $y=(y_i)_{i\in\mathbb A}$, both assumed to be
probability distributions over $\mathbb A$.  If, with the same
conditions on $x$ and $y$, it can be concluded that
$(\pi_i)_{i\in\mathbb A}=(\pi(x_i,y_i))_{i\in\mathbb A}$ is in fact a
probability distribution, just as $x$ and $y$, I will say that $\pi$
is {\it strongly consistent}.

\medskip {\bf 6} To enable {\it observations} from the world, I must {\it
  configure} all available resources such as {\it observation- and
  measuring devices}. The resulting {\it configuration} will enable me
to perform {\it experiments}, i.e. to study individual {\it
  situations} from the world which have my interest.

\medskip {\bf 7} Before actual observations are performed, I must identify the
various possible {\it basic events} (or {\it pure states}, as some may
prefer), which I could encounter. I shall
characterize them by an {\it index}, typically $i$, intended to have
{\it semantic significance.}  The set of possible basic events is the
{\it alphabet} pertaining to the situation, call it $\mathbb A$.  The
actual naming of members of $\mathbb A$, the {\it semiotic assignment},
should catalyze semantic awareness and facilitate technical handling.

\medskip {\bf 8} I will apply a {\it principle of separability} and consider my
{\it total effort} related to observations from the configured
situation to be the sum of individual efforts associated with the
basic events. In so doing, I must weigh each contribution according to
the force with which I will experience the associated basic event. The
total effort I also refer to as {\it total complexity} or simply
{\it complexity} and thus find
that
complexity is the weighted sum of individual complexities:
\begin{equation}\label{eq:cpl}
\Phi(x,y)=\sum_{i\in\mathbb A}\pi(x_i,y_i)\kappa(y_i)\,.
\end{equation}
Here, $x=(x_i)_{i\in\mathbb A}$ and $y=(y_i)_{i\in\mathbb A}$ are,
respectively the truth-assignments and the
belief-assignments associated with the various basic events.

\medskip {\bf 9}
I will attempt to minimize complexity and shall appeal to the
principle that the smallest value for complexity is
obtained when {\it belief matches truth}. As 
\begin{equation}\label{eq:frus}
\sum_{i\in\mathbb A}\pi(x_i,y_i)\kappa(y_i)-\sum_{i\in\mathbb
  A}x_i\kappa(x_i)
\end{equation}
represents my {\it frustration}, the principle says that frustration
is the least, in fact disappears, when $y_i=x_i$ for all $i\in\mathbb
A$.

Given $x=(x_i)_{i\in\mathbb A}$, minimal complexity is what I will aim
at. It is an important quantity. In anticipation, I will call it {\it
  entropy} and denote it by the letter {\it H}\footnote{In order to
  allow a singular case -- the case $q=0$ of Theorem 1 below -- to fit
  into the framework, the infimum in \eqref{eq:ent} should be
  restricted to run over probability distributions $y$ with a support
  which contains the support of $x$.}:
\begin{equation}\label{eq:ent}
\HH(x)=\inf_{y=(y_i)_{i\in\mathbb A}}\Phi(x,y)=\sum_{i\in\mathbb
  A}x_i\kappa(x_i)\,.
\end{equation}
The quantity \eqref{eq:frus} too appears important. It is tempting to
call it {\it ``frustration''} but, again in anticipation,
I better call it {\it divergence}. I shall denote it by the
letter {\it D}:
\begin{equation}\label{eq:div}
\D(x,y)=\Phi(x,y)-\HH(x)\,.
\end{equation}

\section{Conclusion}

\begin{thm} With assumptions and definitions as introduced
  above, assuming only that the interaction is weakly consistent,
 the number
  $q=\pi(1,0)$ must be non-negative and, to each $q\in[0,\infty[$,
  there is only one pair of interaction and coder which fulfill the
  conditions imposed. These functions, denoted $\pi_q$ and $\kappa_q$,
  are determined by the formulas
\begin{align}\label{eq:interact}
\pi_q(x,y)&=qx+(1-q)y\,,\\\label{eq:coder}
\kappa_q(y)&=\ln_q\frac{1}{y}\,,
\end{align}
where the {\it $q$-logarithm} is given by
 \begin{equation}
 \ln_qx=
 \begin{cases}
 \ln x\mbox{ if }q=1,\\
 \frac{x^{1-q}-1}{1-q}\mbox{ if }q\neq 1\,.
 \end{cases}
 \end{equation}
\end{thm}

Note that strong consistency holds if and only if $0\leq q\leq 1$.

The accompanying quantities, complexity, entropy and
divergence are denoted $\Phi_q$, $\HH_q$ and $\D_q$, respectively, and
given through \eqref{eq:cpl}, \eqref{eq:ent} and \eqref{eq:div}, i.e.
\begin{align}\label{eq:cplq}
  \Phi_q(x,y)&=\sum_{i\in\mathbb A}\pi_q(x_i,y_i)\kappa_q(y_i)\,,\\
\label{eq:entq}
\HH_q(x)&=\sum_{i\in\mathbb A}x_i\kappa_q(x_i)\,\\
\label{eq:divq}
\D_q(x,y)&=\sum_{i\in\mathbb A}\Big(\pi_q(x_i,y_i)\kappa_q(y_i)-x_i\kappa_q(x_i)\Big)\,.
\end{align}
In \eqref{eq:entq} we recognize the family of {\it Tsallis
  entropies}, cf. Tsallis \cite{Tsallis88}.  

Regarding the proof of Theorem 1, we shall here only give a brief
indication: The formula \eqref{eq:interact} is readily derived from
the assumption of weak consistency. Then, the only possible form for
the coder, \eqref{eq:coder}, is derived from pretty standard
variational arguments. The final step of the proof, that with
\eqref{eq:interact} and \eqref{eq:coder} the variational principle
does indeed hold, follows by observing the close tie to entropy- and
divergence- measures as derived by an approach due to Bregman, cf. the
recent papers \cite{Topexpo07} and \cite{Naudts08} and references
there.

\section{Hints to the literature} 

Regarding the formula \eqref{eq:entq} for entropy and its
significance, we note that it first appeared in the mathematical
literature in Havrda and Charv{\'a}t \cite{Havrda67},
that it then appeared in the physical literature in Lindhard and
Nielsen \cite{LindhardNielsen71}, and in Lindhard \cite{Lindhard74},
and that it was efficiently promoted in the paper by Tsallis
\cite{Tsallis88} which triggered much research in the physical
community as also witnessed by the many entries in the
database pointed to under \cite{Tsallis88}.

\section{Formal publication}

The present manuscript, posted on the arXiv server, is an
announcement, prior to formal publication. A manuscript with a
comprehensive discussion, with a full proof of Theorem 1 and with some
further results will be worked out soon and submitted to the
electronic journal ``Entropy''. 

Further discussion of the 
considerations presented as well as suggestions of concrete mechanisms
behind the concept of interaction are among obvious issues to look
closer into.

\end{document}